\documentclass{nature}
\usepackage{graphicx}
\usepackage{amsmath}
\usepackage{amsxtra}
\usepackage{amstext}
\usepackage{amssymb}
\usepackage{caption}

\newcommand{\ket}[1]{\ensuremath{|{#1}\rangle}}
\newcommand{\bracket}[2]{\ensuremath{\langle{#1}|{#2}\rangle}}

\title{Quantum acoustics with superconducting qubits}

\author{Yiwen Chu$^1$, Prashanta Kharel$^1$, William H. Renninger$^1$, Luke D. Burkhart$^1$, Luigi Frunzio$^1$, Peter T. Rakich$^1$, \& Robert J. Schoelkopf$^1$}

\begin{document}

\maketitle

\begin{affiliations}
 \item Department  of  Applied  Physics,  Yale  University,  New  Haven,  Connecticut  06511,  USA  and
Yale  Quantum  Institute,  Yale  University,  New  Haven,  Connecticut  06520,  USA
\end{affiliations}

\begin{abstract}
The ability to engineer and manipulate different varieties of quantum mechanical objects allows us to take advantage of their unique properties and create useful hybrid technologies\cite{Kurizki2015}. Thus far, complex quantum states and exquisite quantum control have been demonstrated in systems ranging from trapped ions\cite{Debnath2016, Blatt2012} and solid state qubits\cite{Waldherr2014, Zwanenburg2013} to superconducting microwave resonators\cite{Wang2016, Barends2016}. Recently, there have been many efforts\cite{Aspelmeyer2014, Poot2012} to extend these demonstrations to the motion of complex, macroscopic objects. These mechanical objects have important practical applications in the fields of quantum information and metrology as quantum memories or transducers for measuring and connecting different types of quantum systems. In pursuit of such macroscopic quantum phenomena, mechanical oscillators have been interfaced with quantum devices such as optical cavities and superconducting circuits\cite{Xu2016, Clark2017, Fink2015}. In particular, there have been a few experiments that couple motion to nonlinear quantum objects\cite{OConnell2010, LaHaye2009, Gustafsson2014} such as superconducting qubits. Importantly, this opens up the possibility of creating, storing, and manipulating non-Gaussian quantum states in mechanical degrees of freedom. However, before sophisticated quantum control of mechanical motion can be achieved, we must overcome the challenge of realizing systems with long coherence times while maintaining a sufficient interaction strength. These systems should be implemented in a simple and robust manner that allows for increasing complexity and scalability in the future. Here we experimentally demonstrate a high frequency bulk acoustic wave resonator that is strongly coupled to a superconducting qubit using piezoelectric transduction. In contrast to previous experiments with qubit-mechanical systems\cite{OConnell2010, LaHaye2009, Gustafsson2014}, our device requires only simple fabrication methods, extends coherence times to many microseconds, and provides controllable access to a multitude of phonon modes. We use this system to demonstrate basic quantum operations on the coupled qubit-phonon system. Straightforward improvements to the current device will allow for advanced protocols analogous to what has been shown in optical and microwave resonators, resulting in a novel resource for implementing hybrid quantum technologies.

\end{abstract}

Measuring and controlling the motion of massive objects in the quantum regime is of great interest for both technological applications and for furthering our understanding of quantum mechanics in complex systems. In some respects, the physics of phonons inside a crystal is similar to that of photons inside an electromagnetic resonator, which are routinely treated as quantum mechanical objects. However, such mechanical excitations involve the collective motion of a large number of atoms in the complex environment of a macroscopic object. Therefore, it remains an open question whether mechanical objects can be engineered, controlled, and utilized in ways analogous to what has been demonstrated in cavity\cite{ReisererRMP2015} or circuit QED\cite{Wang2016}. By addressing this question, we may be able to use mechanical systems as powerful resources for quantum information and metrology, such as universal transducers or quantum memories that are more compact than their electromagnetic counterparts\cite{Schuetz2015, Aspelmeyer2014, Poot2012, Bochmann2013}. In addition, since any coupling of qubits to other degrees of freedom can lead to decoherence, it is crucial to understand and control the interaction qubits might have to their mechanical environments\cite{IoffePRL2004}. 

In the field of quantum electro-mechanics, there has been a variety of experimental efforts to couple mechanical motion to superconducting circuits.
The majority of demonstrations have involved megahertz frequency micromechanical oscillators parametrically coupled to gigahertz frequency electromagnetic resonators\cite{Clark2017, Fink2015}. Because both electrical and mechanical modes are linear, these systems only allows for the generation of Gaussian states of mechanical motion. Alternatively, the creation of useful non-Gaussian states, including Fock states or Schr\"{o}dinger cat states, requires introducing a source of quantum nonlinearity, such as a qubit. 

Demonstrations of mechanics coupled to superconducting qubits include interactions with propagating surface acoustic waves\cite{Gustafsson2014} and micromechanical resonators in both the dispersive\cite{LaHaye2009, RouxinolNanotech2016} and resonant\cite{OConnell2010} regimes. A central goal of these experiments is to reach the regime of quantum acoustics, in which the ability to make, manipulate, and measure non-classical states of light in cavity or circuit QED becomes applicable to mechanical degrees of freedom. This regime requires the strong coupling limit, where the coupling strength $g$ is much larger than the loss rates $\gamma, \kappa$ of both the qubit and the oscillator. Piezoelectric materials are natural choices for achieving large coupling strengths between single electrical and mechanical excitations\cite{Han2016}. These coupling strengths can be as large as tens of MHz, comparable to qubit-photon couplings possible in circuit QED. Nevertheless, there has only been one demonstration of a nonlinear electromechanical system in the strong coupling limit\cite{OConnell2010}. The outstanding question is how to simultaneously achieve coherences and coupling strengths that would allow for sophisticated quantum operations in a robust and easily implemented electromechanical system.



In this Letter, we address this question by experimentally demonstrating strong coupling between a superconducting qubit and the phonon modes of a high-overtone bulk acoustic wave resonator (HBAR). The system involves the straightforward incorporation of a piezoelectric transducer into a standard 3D transmon geometry\cite{Paik:2011hd}. The relatively simple fabrication procedure achieves a qubit coherence of many microseconds and enables more advanced geometries in the future. We are able to individually address many acoustic modes in our system. By performing basic quantum operations with the qubit, we reach the mechanical ground state and observe long coherence times ($>$10 $\mu$s) of single phonons. The cooperativity $C=g^2/\kappa\gamma$ of our system is 260, comparable to that of early circuit QED devices\cite{Wallraff2004} and more than an order of magnitude higher than previous qubit coupled mechanical systems\cite{OConnell2010}.

Our quantum electromechanical device consists of a frequency-tunable aluminum transmon qubit coupled to phonons in its non-piezoelectric sapphire substrate using a thin disk of c-axis oriented aluminum nitride (AlN) (Figure 1a). We pattern the disk from a commercially purchased wafer of AlN film on sapphire and fabricate the qubit directly on top using standard techniques (see Supplementary Information). The substrate surfaces form a phononic Fabry-Perot resonator that supports longitudinally polarized thickness modes (see Figure 1b), which are well studied in the context of conventional HBAR technologies such as quartz resonators\cite{ZhangJAP2006}. The piezoelectricity of the AlN generates stress $\overset\leftrightarrow{\sigma}(\vec{x})$ from the transmon's electric field $\vec{E}(\vec{x})$, which acts on the phonon modes's strain field $\overset\leftrightarrow{s}(\vec{x})$. For simplicity, we consider only the dominant tensor components $E \equiv E_3$, $\sigma \equiv \sigma_{33}$,  $s \equiv s_{33}$, where the subscript $3$ denotes the longitudinal direction perpendicular to the substrate surface. Then the interaction energy between the qubit and the phonon mode is given by $H = -\int{\sigma(\vec{x}) s(\vec{x})}\: dV$, where $\sigma(\vec{x}) = c_{33} d_{33}(\vec{x}) E(\vec{x})$ and $c_{33}$ and $d_{33}$ are the stiffness and piezoelectric tensor components, respectively. Quantizing the fields and equating this to the Jaynes-Cummings Hamiltonian $H_{\textrm{int}} = -\hbar g(ab^\dagger+a^\dagger b)$, where $a$ and $b$ are the annihilation operators for the qubit and phonon modes, respectively, we can estimate the coupling strength as $\hbar g = c_{33}\int{d_{33}(\vec{x})E(\vec{x}) s(\vec{x})}\: dV$ (see Supplementary Information for details).

Having described the physics of the electromechanical coupling, we now introduce a simple picture that captures the essential character of the acoustic modes and allows us to estimate coupling rates and mode frequencies. Because the acoustic wavelength is much smaller than the diameter of the AlN disk, the transduced acoustic waves do not diffract significantly and remain inside the cylindrical volume of sapphire underneath the AlN for a relatively long time. As shown in Figure 1b, the spatial character and frequencies of the phonons can be approximated by considering the stationary modes of this cylindrical mode volume, which have strain field distributions given by
\begin{equation}
s_{l, m}(\vec{x}) = \beta_{l, m}\textrm{sin}\left(\frac{l\pi z}{h}\right)J_0\left(\frac{2j_{0,m}r}{d} \right),
\label{eq:mode}
\end{equation}
where $J_0$ is the zeroth order Bessel function of the first kind and $j_{0,m}$ is the $m^{\textrm{th}}$ root of $J_0$. As indicated in Figure 1b, $h$ is the substrate thickness and $d$ is the disk diameter. $\beta_{l, m}$ is a normalization factor corresponding to the root-mean-squared strain amplitude of a single phonon at frequency
\begin{equation}
\omega_{l, m} = \sqrt{\left(\frac{l\pi }{h}\right)^2 v_l^2+\left(\frac{2j_{0,m}}{d}\right)^2 v_t^2 }.
\label{eq:wnm}
\end{equation}
Here $v_l$ and $v_t$ are the longitudinal and effective transverse sound velocities, respectively. According to this simplified model, we describe the qubit as coupling to discrete modes with distinct longitudinal ($l$) and transverse ($m$) mode numbers. For example, the $l = 503, m = 0$ phonon mode has a frequency of $\sim$6.65 GHz. We can obtain $E(\vec{x})$ from electromagnetic simulations of a qubit at that frequency and estimate the coupling strength $g$ to be on the order of $2\pi \times 300$ kHz. 

In order to reach the strong coupling limit, $g$ needs to be much larger than the mechanical loss, which we expect to be dominated by diffraction out of the finite mode volume into the semi-infinite sapphire substrate. To estimate this loss, we consider a second model that better approximates the actual physical system with a large cylindrical substrate with diameter $a\gg d$. Now the qubit can be seen as coupling to an almost continuous set of lossless modes $s'_{l, m}(\vec{x})$, which are also described by Equations \ref{eq:mode} and \ref{eq:wnm}, except with $d$ replaced by $a$. The coherent temporal evolution of these eigenmodes will conspire to reproduce the diffraction loss of the original strain profile within a timescale $\sim a/v_t$. As shown in the Supplementary Information, we use this method to estimate the phonon's diffraction limited lifetime to be on the order of many microseconds. The estimated lifetime confirms the validity of using a simpler model of discrete modes with high quality factors and indicates that our system should be in the strong coupling regime. 

We see from these descriptions that the modes of our mechanical system are physically very different from the micromechanical resonators used in other works\cite{LaHaye2009, OConnell2010}. Even though the mechanical excitations are not physically confined in all directions, they can nevertheless be described using the well defined modes of a finite volume. We will show experimentally that, in accordance with theoretical estimates, this geometric loss still results in a much higher quality factor than previously demonstrated micromechanical resonators at the same frequency\cite{OConnell2010}. In addition, a greater fraction of the mechanical energy in our system resides in an almost perfect crystal rather than in potentially lossy interfaces and surfaces\cite{Poot2012}. Combined with the lack of complex fabrication processes that could further increase material dissipation, we expect our system to be a path toward very low loss mechanical resonators, in analogy to the case of long-lived 3D electromagnetic resonators\cite{Reagor:2013tq}. 

We now turn to experiments that explore the physics of the coupled qubit-phonon system. The mechanically coupled qubit is placed inside a copper rectangular 3D cavity at $\nu_c$ = 9.16 GHz with externally attached flux tuning coils. This device is mounted on the base plate of a dilution refrigerator and measured using standard dispersive readout techniques with a Josephson parametric converter amplifier\cite{Abdo2011}.

By performing spectroscopy on the transmon qubit, we are able to observe the hallmarks of strong coupling to the modes of the HBAR. First, as we perform saturation spectroscopy on the qubit while varying its frequency with applied flux, we observe a series of evenly spaced anticrossings, which are consistent with phonons of different longitudinal mode numbers (Figure 2a). These features occur every $\nu_{\textrm{FSR}}$ = $v_l/2h$ =13.2 MHz as we tune the transmon's frequency by over a gigahertz (see Supplementary Information). For a measured substrate thickness of 420 $\mu$m, $\nu_{\textrm{FSR}}$ corresponds to the free spectral range of a HBAR with longitudinal sound velocity $v_l$=1.11$\times$10$^4$ m/s, which agrees well with previously measured values for sapphire.
 Finer features appear in more detailed spectroscopy data, shown in Figure 2b and the inset to Figure 2a. We also observe corresponding features in qubit $T_1$ lifetime measurements (Figure 2c). Far away from the anticrossing point, we measure an exponential decay of the qubit excited state population with $T_1=6$ $\mu$s. Around the anticrossing, we observe clear evidence of vacuum Rabi oscillations. The oscillations are distorted on the lower current (higher qubit frequency) side of the main feature, and there are regions of what appears to be lower qubit $T_1$'s. These fine features reproduce for all longitudinal modes we investigated, though their shape and relative spacings change gradually with frequency. The frequencies of these features indicate that they correspond to the modes $s_{l, m}(\vec{x})$ with the same $l$ and different $m$. By simulating the experiments in Figure 2b and c using the first four transverse mode numbers (see Supplementary Information), we find good agreement with the data and extract a coupling constant for the $m = 0$ mode of $g = 2\pi\times (260 \pm 10)$ kHz, which agrees reasonably well with our prediction of $2\pi \times 300$ kHz.

Now that we have investigated and understood the electromechanical coupling in our device, we show that it can be used to perform coherent manipulations and quantum operations on the qubit-phonon system.
The qubit's interaction with each phonon mode can be effectively turned on and off by tuning it on and off resonance with that mode. To perform useful quantum operations this way, the tuning must be performed over a frequency range much larger than $g$ and on a timescale much faster than one vacuum Rabi oscillation period. This is difficult to achieve using flux tuning, but can be easily accomplished by Stark shifting the qubit with an additional microwave drive\cite{Leghtas2015}. We apply a constant flux so that the qubit is at the frequency $\omega_\textrm{b}$ indicated in Figure 2b. In order to avoid coupling to the higher order transverse modes, we apply a drive 100 MHz detuned from the microwave cavity frequency $\nu_c$ with a constant amplitude that Stark shifts the qubit to $\omega_{\textrm{or}}$. This is the off resonant frequency of the qubit where it is not interacting with any phonons and can be controlled and measured as an uncoupled transmon. Decreasing the Stark shift amplitude makes the interaction resonant, allowing for the exchange of energy between the qubit and phonon. 

To calibrate the Stark shift control, we reproduce the vacuum Rabi oscillations shown in Figure 2c using a pulsed Stark drive. From this data, we can determine an amplitude and length of the pulse, indicated by a white cross in Figure 3a, that constitutes a swap operation between the qubit and phonon. This swap operation allows us to transfer a single electromagnetic excitation of the nonlinear transmon into a mechanical excitation of the phonon and vice versa. 

We first use our ability to perform operations on the qubit-phonon system to show that the mechanical oscillator is in the quantum ground state. By using a protocol that measures the amplitude of Rabi oscillations between the e and f states of the qubit\cite{Geerlings2013}, we find that it has a ground state population of $92\%$ (Figure 3b). Ideally, the qubit and phonon should be in their ground states since both are in the regime of $k_BT<<\hbar\omega$. If we first perform a swap operation between the qubit and phonon, we find that the qubit's ground state population increases to $98\%$. This value is likely limited by the fidelity of the swap operation therefore represents a lower bound on the phonon ground state population. This result indicates that that the phonons are indeed cooled to the quantum ground state, in fact more so than the transmon qubit. The swap operation can be used to increase the qubit polarization with the phonon mode, which can also be seen in an increased contrast of g-e Rabi oscillations (Figure 3c).

To further verify that our system is indeed in the strong coupling regime, we now present measurements of the phonon's coherence properties. To measure the phonon $T_1$, we excite the qubit and then perform a swap operation, thus preparing the phonon in the $n=1$ Fock state (Figure 4a). We then perform another swap after a variable delay and measure the qubit population in the excited state. We find that the data is well described by an exponential decay with a time constant of $T_1$ = $17\pm 1$ $\mu$s with the addition of a decaying sinusoid with frequency of $2\pi\times (340 \pm 10)$ kHz, which corresponds approximately to the frequency difference between the $m=0$ and $m=1$ modes. 
 We also measure a phonon $T_2$ decoherence time of $27 \pm 1$ $\mu$s using a modified Ramsey sequence (Figure 4b). In the Supplementary Information, we present additional data showing $T_1$ measurements where an ``imperfect" swap pulse is used, resulting in different decay lifetimes and quantum interference between the qubit and phonon states. 


The results presented here are the first demonstrations of an electromechanical system with significant room for improvement. We have already shown that with simple modifications to a standard 3D transmon geometry, we can perform quantum operations on a highly coherent mechanical mode. There are clear paths toward enhancing both the coherence and interaction strength of the system to bring it further into the strong coupling regime. The most obvious improvement is to increase the transmon $T_1$, which is currently the limiting lifetime in the system. 3D transmons with $T_1\sim$100 $\mu$s have been demonstrated on sapphire. It is unclear if AlN currently causes additional dielectric loss, but tan$\delta\sim10^{-3}$ have been measured for AlN\cite{OConnellAPL2008}. Given the simulated dielectric participation ratio of our transducer, this places a limit on the qubit $T_1$ of $\sim$1 ms. Another significant improvement would be to modify the geometry so that the qubit couples more strongly to a single, long-lived phonon mode. This can be done by shaping the surfaces of the substrate to create a stable phonon resonator with transverse confinement\cite{Renninger2016, Galliou2013}. The AlN transducer can also be made with a curved profile to minimize higher spatial Fourier components of the piezoelectric drive\cite{Park2001}. 

These improvements will open up possibilities for more sophisticated quantum acoustics demonstrations in the future. With stronger coupling and lower loss, we can treat the phonons analogously to the modes of an electromagnetic resonator. For example, with tools that we have already demonstrated, we will be able to create and read out higher phonon Fock states\cite{Hofheinz2008}. There is also evidence that in our system, the phonons can be directly displaced into a coherent state with a microwave drive, and it may be possible to reach the strong dispersive regime for the qubit-phonon interaction. The combination of these abilities will allow us to create highly non-classical mechanical states such as Schrodinger cat states\cite{Hofheinz2009}. In addition, phonons may offer distinct advantages over photons as a quantum resource in cQED devices. Large quality factors of up to $\sim10^8$ have been demonstrated in bulk acoustic waves resonators\cite{Galliou2013, Renninger2016, Han2016}, which is comparable to the longest lived 3D superconducting cavities. However, the phonons can be confined to a much smaller mode volume due to the difference in the speed of sound and light. Furthermore, this small volume supports a large number of orthogonal longitudinal modes that can all be coupled to the qubit, resulting in a multimode register for the storage of quantum information. In addition, our results indicate that phonon radiation could be a loss mechanism for superconducting circuits if piezoelectric materials are present\cite{Dai2011, IoffePRL2004}. Finally, bulk acoustic waves have been shown to couple to a variety of other quantum mechanical objects ranging from infrared photons to solid state defects\cite{Renninger2016, MacQuarrie2013}. Therefore, our device presents new possibilities for microwave to optical conversion and transduction in hybrid quantum systems. 

 \newpage

\begin{center}
\includegraphics[width = 0.5\textwidth]{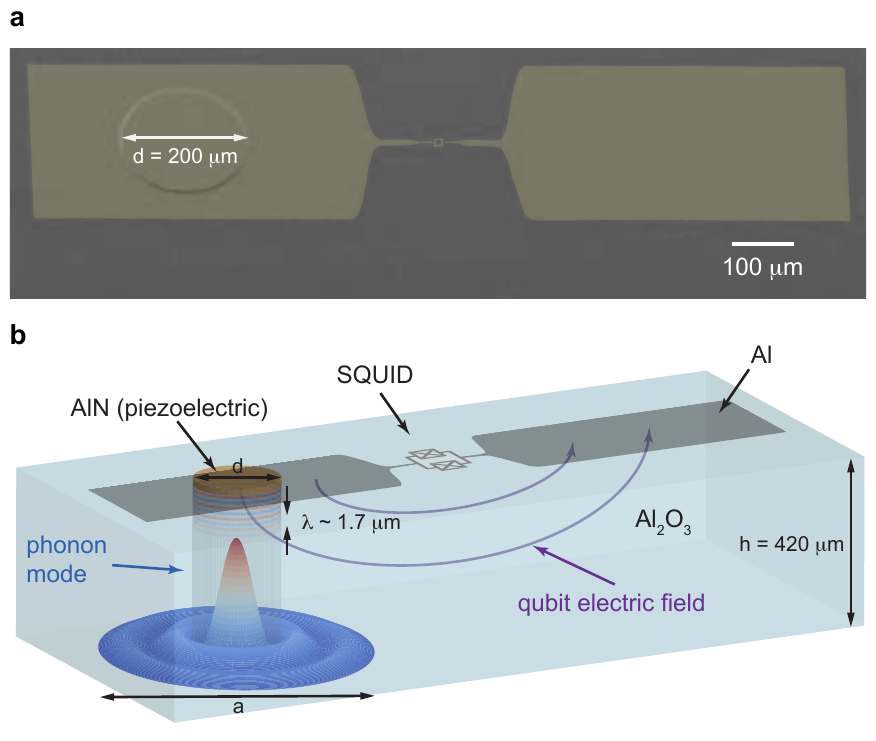}
\captionof{figure}{
\textbf{Qubit with piezoelectric transducer. a}, False color SEM image of a transmon qubit on a sapphire substrate with one electrode covering an AlN transducer, which is $\sim$ 900 nm thick and $d$ = 200 $\mu$m in diameter. \textbf{b}. Schematic of piezoelectric coupling to the modes of a HBAR (not to scale). 
The longitudinal part of the wavefunction given in Equation \ref{eq:mode} is illustrated by a sinusoidal profile with wavelength $\lambda = 2h/l$ on the cylindrical mode volume defined by the transducer. The transverse energy density profile of $s_{l, 0}(\vec{x})$ is plotted in 3D, showing the effective confinement of energy inside the mode volume, while some energy leaks out due to diffraction. This also illustrates that the $s_{l, 0}(\vec{x})$ mode is equivalent to the $s'_{l, 3}(\vec{x})$ mode of a larger volume with diameter $a$.
}
\end{center}

 \newpage

\begin{center}
\includegraphics[width = \textwidth]{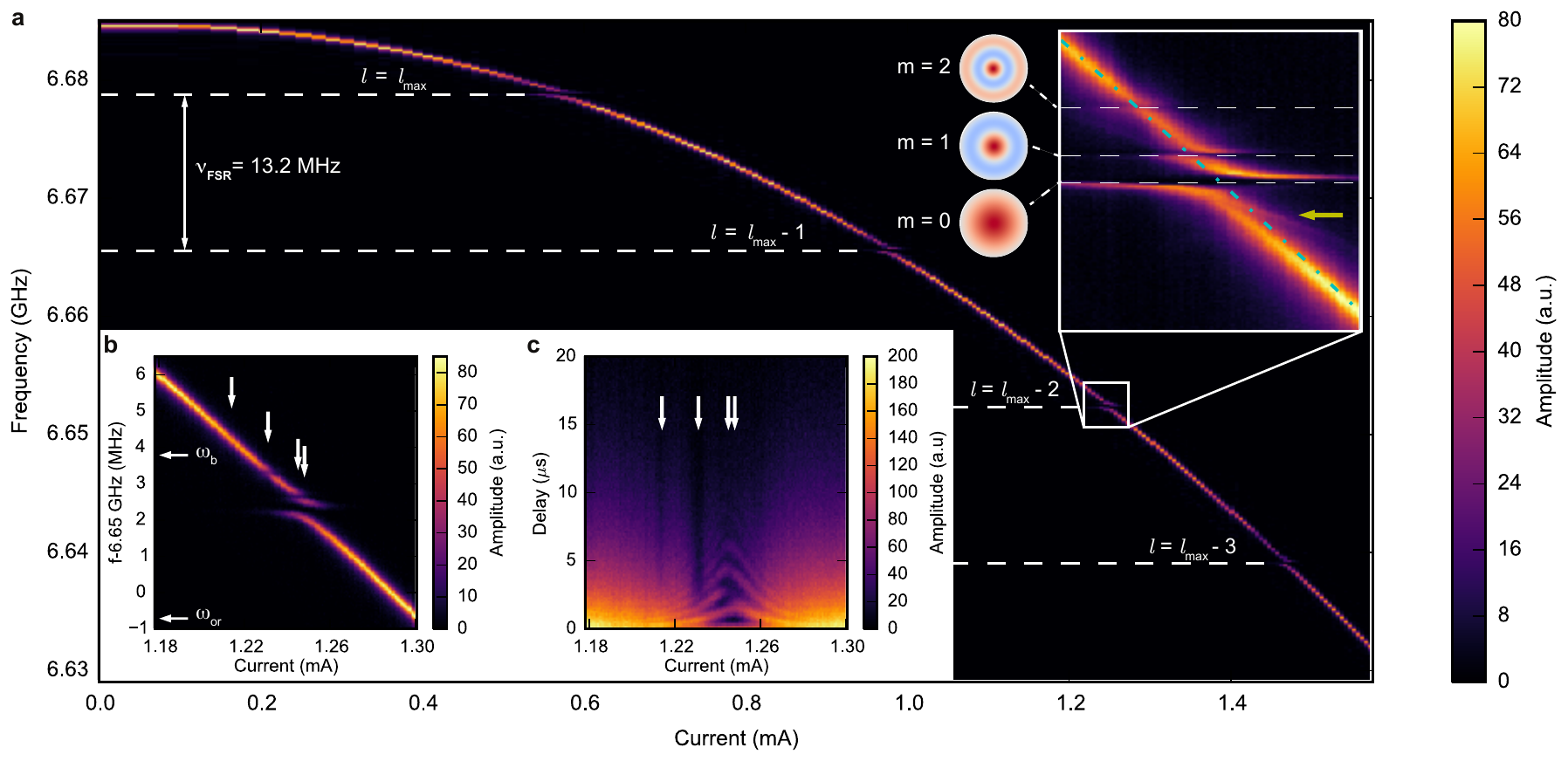}
\captionof{figure}{\textbf{Spectroscopy of qubit-phonon coupling. a}, Qubit spectroscopy as a function of current applied to flux tuning coil. 
White dashed lines indicate the locations of anticrossings for different longitudinal wavenumbers. The highest accessible longitudinal mode is $l_{max} = 505$, assuming $v_l$ is constant with frequency. Inset: Detailed spectroscopy around the $l=503$ anticrossing, which is also used in b and c, along with Figures 3 and 4. Blue dash-dot line shows the frequency of the uncoupled qubit. Dashed white lines show the locations of anticrossings for $m = 0, 1, 2$, whose transverse mode profiles are plotted to the left. The faint feature indicated by a yellow arrow is due to multiphoton transitions to higher states of the Jaynes-Cummings level structure\cite{Bishop2008}. \textbf{b, c}, Spectroscopy and qubit $T_1$ measurements. Vertical arrows indicate locations of prominent subfeatures. Horizontal arrows in b indicate frequencies used for Stark shift control, as described in the text. The $T_1$ measurement was performed using a $20$ ns microwave pulse at the qubit frequency followed by qubit state readout after a variable delay. Near the anticrossing, where the qubit frequency is not well defined, the large pulse bandwidth ensures that the qubit component of the hybridized states are excited.}
\end{center}
 \newpage
 
\begin{center}
\includegraphics[width = 0.5\textwidth]{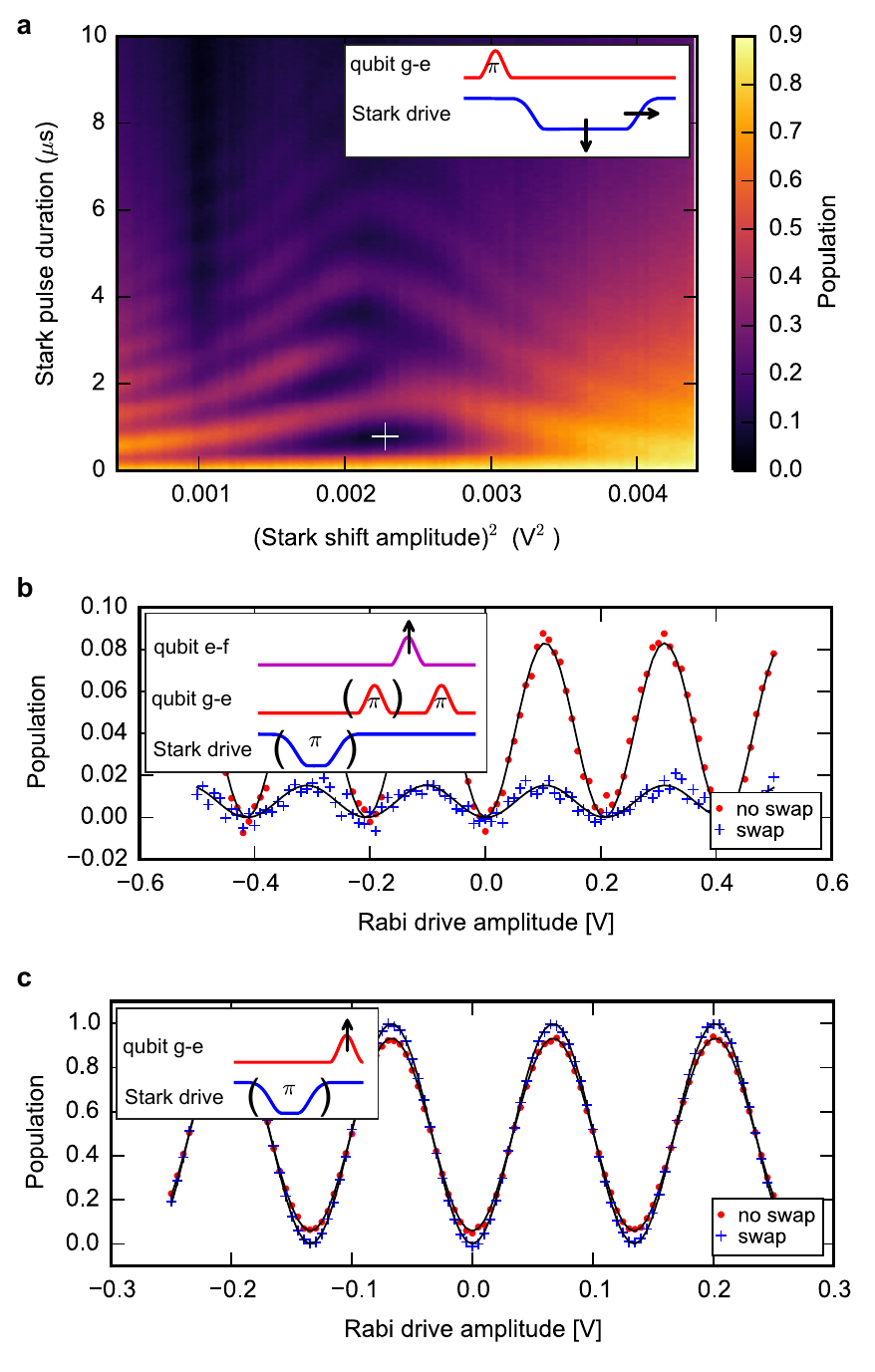}
\captionof{figure}{\textbf{Quantum control of the qubit-phonon system. a}, Vacuum Rabi oscillations measured by varying the amplitude and duration of the Stark drive pulse after exciting the qubit while it is off resonant from the phonons, as shown in the inset. The pulse is a decrease in the Stark drive amplitude with a rise time of 50 ns, which is faster than $1/g$ but has a bandwidth less than the drive detuning to avoid populating the microwave cavity. White cross shows the pulse amplitude and duration for an optimal swap operation between the qubit and phonon. Axes labeled "Population" Figures 3 and 4 correspond to all populations not in the g state.
\textbf{b}, Measurement of the excited state populations of the qubit and phonon. In both cases, we first drive Rabi oscillations between the qubit's e and f states followed by a g-e $\pi$ pulse\cite{Geerlings2013}. This experiment is then repeated with a preceding g-e $\pi$ pulse. We plot the Rabi oscillations obtained in the first experiment normalized by the sum of the fitted oscillation amplitudes from both experiments. The amplitude of the normalized oscillations gives the population in the $n=1$ Fock state of the phonon or the e state of the qubit, depending on whether or not a swap operation is performed at the beginning. Black lines show sinusoidal fits to the data.
\textbf{c}, Rabi oscillations between the g and e qubit states, with and without a preceding swap operation. We use the former to calibrate the qubit population measurements in Figures 3 and 4.
}
\end{center}
 \newpage
\begin{center}
\includegraphics[width = 0.5 \textwidth]{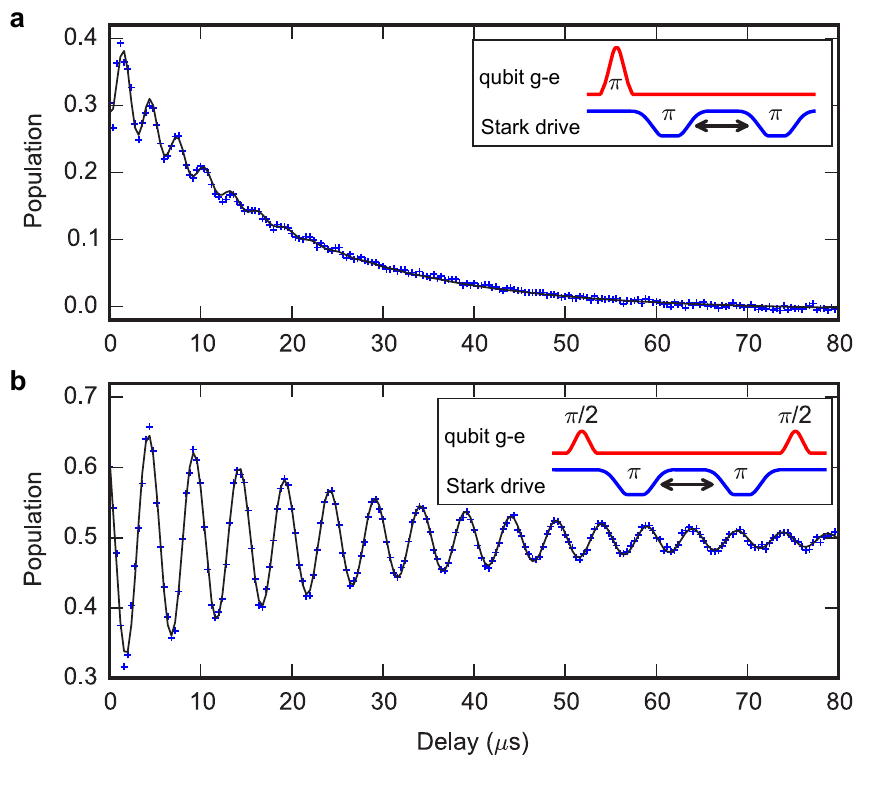}
\captionof{figure}{\textbf{Phonon coherence properties. a}, Phonon $T_1$ measurement. Black line is a fit to an exponential decay plus a decaying sinusoid.
\textbf{b}, Phonon $T_2$ measurement. The phase of the second $\pi/2$ pulse is set to be $(\omega_0+\Omega)t$, where $t$ is the delay, $\omega_0$ is the detuning between the qubit and phonon during the delay, and $\Omega$ provides an additional artificial detuning. Black line is a fit to an exponentially decaying sinusoid with frequency $\Omega$.
}
\end{center}

 \newpage
%
%
%


\bibliography{QubitPhononArxiv.bib}


\begin{addendum}
 \item We thank Michel Devoret, Konrad Lehnert, Hong Tang, and Hojoong Jung for helpful discussions. We thank Katrina Silwa for providing the Josephson parametric converter amplifier. This research was supported by the US Army Research Office (W911NF-14-1-0011). Facilities use was supported by the Yale SEAS cleanroom, the Yale Institute for Nanoscience and Quantum Engineering (YINQE), and the NSF (MRSEC DMR 1119826). L.D.B acknowledges support from the ARO QuaCGR Fellowship.
 \item[Correspondence] Correspondence and requests for materials
should be addressed to Y. Chu (email: yiwen.chu@yale.edu) or R. J. Schoelkopf~(robert.schoelkopf@yale.edu ).
\end{addendum}

\clearpage


\pagebreak
\begin{center}
\textbf{\large Supplementary information for:\\ Quantum acoustics with superconducting qubits}
\end{center}

\setcounter{equation}{0}
\setcounter{figure}{0}
\setcounter{table}{0}
\setcounter{page}{1}
\setcounter{section}{0}
\makeatletter
\renewcommand{\theequation}{S\arabic{equation}}
\renewcommand{\thefigure}{S\arabic{figure}}
\renewcommand{\thetable}{S\arabic{table}}

\section{Fabrication procedures}
\begin{figure}[ht]
\begin{center}
\includegraphics[width = 0.9 \textwidth]{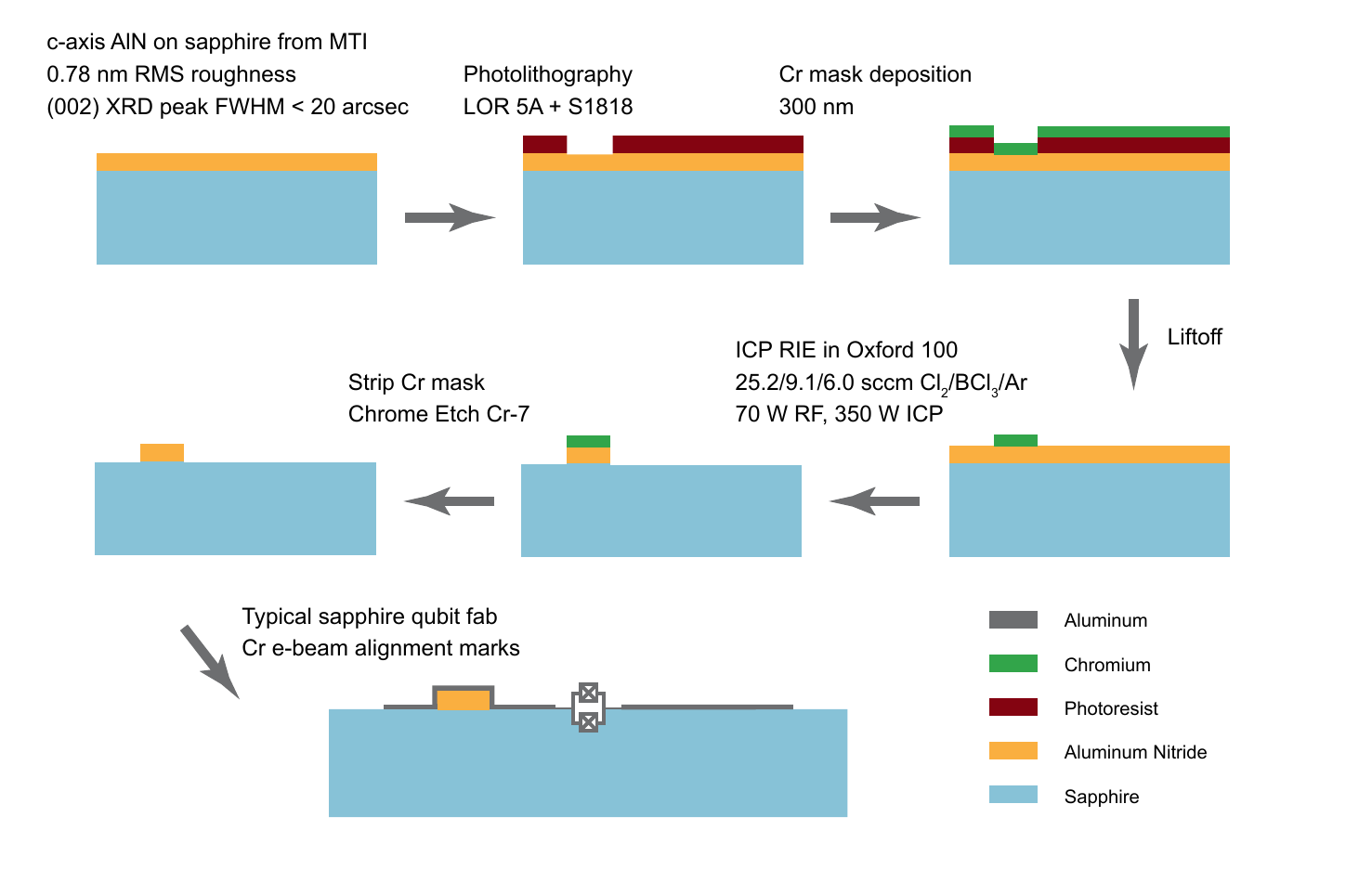}
\caption{\textbf{Fabrication procedure.} Drawings show cross sectional cuts lengthwise through the center of the transmon qubit. Drawing is not to scale. Not included is a photolithography step before RIE etching that masks the Cr alignment markers.}
\label{fab}
\end{center}
\end{figure}

As shown in Figure \ref{fab}, we begin with commercially purchased double side polished 2" sapphire wafers with a 1$\mu$m thick film of c-axis oriented AlN grown on one side (MTI Corporation, part number AN-AT-50-U-1000-C2-US). We use a bilayer of photoresist (LOR 5A, 4000 rpm and S1818, 4000 rpm) for liftoff patterning of a 300 nm thick e-beam deposited chromium hard mask. After liftoff, circular regions of the AlN are masked by Cr, along with alignment markers for e-beam lithography. Another layer of photoresist (S1818, 4000 rpm) is patterned to mask off the regions with alignment markers so that they are protected during the the subsequent steps. We then perform a reactive ion etch (RIE) in an Oxford 100 etcher to define the AlN disks (Cl$_2$/BCl$_3$/Ar at 4/26/10 sccm, 8 mTorr, 70 W RF power, 350 W ICP power). Since the photoresist protecting the alignment marks is not very robust against this etch, we also physically mask off the alignment marks with silicon wafer chips. The Cr masking the AlN disks is removed using a wet etch (Cyantek, Cr-7). The wafer is then placed back in the etcher to thin the AlN to $\sim$ 900 nm. The sapphire is also etched slightly with this RIE chemistry, but with a slower etch rate than the AlN. We slightly over-etch the AlN to make sure that there is no AlN remaining on the rest of the substrate away from the transducer disks.

The transmon qubit is fabricated using a standard e-beam lithography and Dolan bridge evaporation technique\cite{SWang2015}. The Cr markers were used to align the qubit to the AlN disk. The deposited aluminum covers edges of the AlN so that the top of the disk is electrically connected to the rest of the electrode.

\section{Frequency dependence of phonon modes}
\begin{figure}
\begin{center}
\includegraphics[width = 0.9 \textwidth]{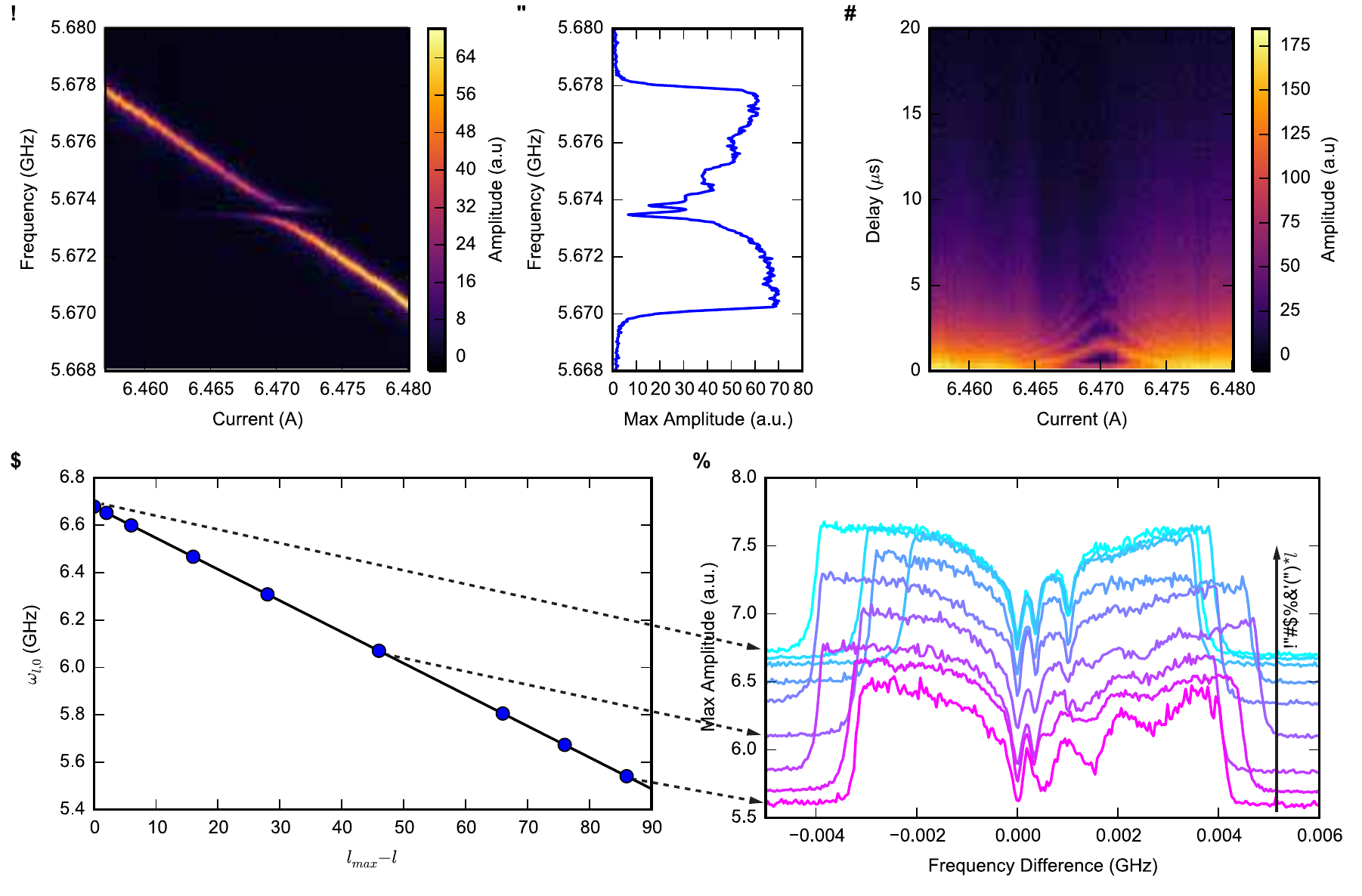}
\caption{\textbf{Phonon modes at different frequencies. a}, Spectroscopy around the $l=429$ mode. \textbf{b}, Maximum values from each horizontal line of \textbf{a}. From here on and in the text, we will call this a maximum value plot. \textbf{c}, T$_1$ and vacuum Rabi oscillations obtained using the same method as Figure 2c in the main text. \textbf{d}, Frequency of the $m=0$ transverse mode $\omega_{l, 0}$ versus the longitudinal mode number. The black line is a linear fit to the data. \textbf{e}, Maximum value plots for the longitudinal modes shown in \textbf{d} after subtracting the frequency of the $m=0$ transverse mode for each. The plots have been shifted vertically so that the background level for each curve approximately corresponds to $\omega_{l, 0}$ in gigahertz on the vertical axis.}
\label{FSRs}
\end{center}
\end{figure}

As mentioned in the main text, we tuned the qubit frequency by more than a gigahertz and observed anticrossing features every 13.2 MHz, corresponding to the longitudinal free spectral range (FSR) of the phonon Fabry-Perot resonator. In Figure 2c and d of the main text, we presented spectroscopy and vacuum Rabi oscillations around the $l = l_{max} -2 = 503$ longitudinal mode number. In Figure \ref{FSRs}a and c, we show results of the same experiments around $l = l_{max} -76 = 429$. In order to condense the information in Figure \ref{FSRs}a, we take the maximum value in each horizontal line and plot them in \ref{FSRs}b. This maximum value plot highlights the substructure, including a dominant anticrossing and a series of other features at higher frequencies. It is apparent in both the spectroscopy and vacuum Rabi data that the substructure is qualitatively different from that of the $l = 503$ data in Figure 2 of the main text.

We investigated nine different FSR's to study this phenomenon. The frequency of the main anticrossing feature, which corresponds to the deepest dip in the maximum value plots, is plotted in \ref{FSRs}d as function of FSR number. A linear fit gives a longitudinal phonon velocity of $v_l$=1.11$\times$10$^4$ m/s. In Figure \ref{FSRs}e, we show the maximum value plots for each longitudinal mode number. According to Equation 2 in the main text, different values of $l$ should only give rise to an overall scaling of the relative frequencies of the transverse modes. We see, however, that the observed behavior is more complex, with the relative locations and intensities of the subfeatures changing with $l$. This could be explained by the presence of the AlN, which provides some frequency-dependent confinement for the phonons in the transverse directions.

\section{Phonon $T_1$ versus swap operation parameters}
\begin{figure}
\begin{center}
\includegraphics[width = 0.9 \textwidth]{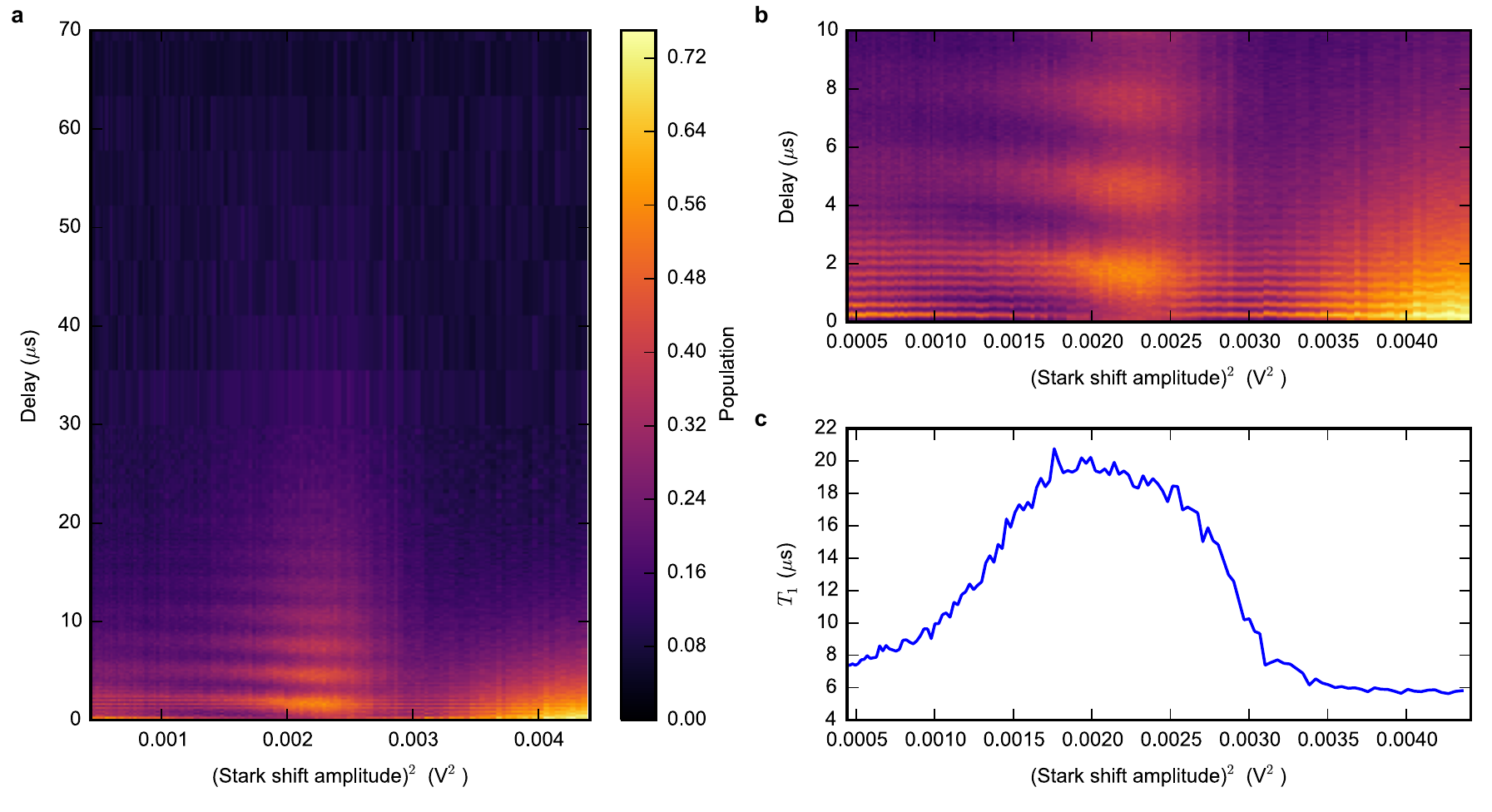}
\caption{\textbf{$T_1$ measurements for different qubit detunings. a}, For each Stark pulse amplitude, we set the "swap" pulse length to be the location of the first local minimum in the vacuum Rabi data in Figure 3a of the main text and perform a $T_1$ measurement using the pulse sequence shown in Figure 4a of the main text. The results are plotted along the vertical axis. \textbf{b}, Detailed view of \textbf{a} to show fast oscillations at short times. \textbf{c}. Effective $T_1$'s obtained by an exponential fit to the data for each Stark pulse amplitude.}
\label{fig:T1s}
\end{center}
\end{figure}

In Figure 4a of the main text, we presented phonon $T_1$ data taken using an optimal swap pulse between the phonon and qubit, whose length and amplitude are indicated by the white cross in Figure 3a of the main text. However, since there is more than one phonon mode with different transverse mode numbers, frequencies, and coupling strengths to the qubit, we investigated how changing the swap pulse length and amplitude affects the measurement of phonon $T_1$. The results are presented in Figure \ref{fig:T1s}. In Figure \ref{fig:T1s}a and b, we see that, in addition to the slow oscillations corresponding to the frequency difference between the $m=0$ and $m=1$ modes, there are fast oscillations with a frequency of $\sim$ 3 MHz. This corresponds to the detuning between the qubit during the swap pulse and when it is tuned away from the phonons during the delay. When the swap pulse is imperfect, some population remains in the qubit during the delay period and evolves at this detuning relative to the phonon. When the qubit is brought back into resonance with the phonon for the second swap pulse, the populations interfere and give rise to the oscillations seen in the data. 

In Figure \ref{fig:T1s}c, we show the effective $T_1$ for each Stark pulse amplitude. We see that the $T_1$ is maximum when the qubit population is most efficiently swapped into the phonon. Away from these strongly coupled phonon modes, the $T_1$ gradually reverts back to that of the uncoupled qubit. The data in Figure \ref{fig:T1s} clearly shows that we are performing coherent operations between the qubit and phonon, and that, in this case, the phonon can actually be used as a quantum memory that is longer lived than the qubit.

\section{Analysis of phonon modes}
In the main text, we presented two complimentary ways to describe the mode structure and geometric loss mechanisms of our system. Here, we explain these conceptual pictures in more detail and give examples of how each can be used to analyze and understand different aspects of the system. These descriptions are intended to give some physical intuition and qualitatively reproduce the experimental results. We then present in the next section numerical simulations of a more exact reproduction of the actual geometry.

Typically in cavity or circuit QED, we imagine an atom or superconducting qubit coupled to a resonator, which is in turn coupled to the outside world through a partially reflecting mirror or an evanescently coupled transmission line. We can describe the resonator as a photon box with well-defined modes, and simplify the coupling to the modes of the surrounding environment into a loss rate for each mode. There is, however, the opposite limit where the cavity is eliminated, such as an antenna or an atom radiating into free space. In the case of the atom, for example, the Wigner-Weisskopf theory of the spontaneous emission rate involves putting the atom in a large fictitious box and considering its coupling to the semi-continuum modes of this box.

Physically, our system has features that resemble both a phonon box and a piezoelectric transducer radiating into the ``free space" of a bulk crystal. On the one hand, except for the 0.9 $\mu$m disk of AlN protruding from one surface of the 420 $\mu$m thick substrate, there are no structures in the transverse direction to confine the phonons. It is a phonon box with two ends but almost no walls. The phonons are essentially free to propagate and diffract out of the box in the transverse directions, which are semi-infinite due to the lossy boundaries of the sapphire. Therefore, as in the case of the radiating atom, it seems reasonable to use the of modes of a semi-inifinite system to calculate the diffraction loss of the phonons. On the other hand, if one considers the strain field produced by the AlN transducer, one realizes that no transverse confinement is required to maintain its energy for a long time inside the cylinder of substrate beneath the AlN. In other words, if we consider this cylinder as our fictitious box, the strain field resembles a mode of this box with a high quality factor. This is because the diameter of the AlN is $\sim$ 100 times larger than the phonon wavelength. Therefore, the phonon mode undergoes very little diffraction as it propagates in the longitudinal direction. 

In order to model the phonon modes and their loss due to diffraction, we will use techniques and concepts from both the lossy resonator and free space radiation pictures. We point out that, in fact, any system with geometric or radiation loss can be simultaneously described in either picture, depending on where one choses the boundaries of the system to be. In most cases, one of the pictures is simply much more convenient for accurately modeling a particular aspect of the system behavior.


\subsection{Vacuum Rabi oscillations}
We first show two different ways of simulating the vacuum Rabi oscillations shown in Figure \ref{FSRs}c and Figure 2c of the main text. The first method uses the modes of a fictitious cylindrical box below the AlN, whose strain profiles are given in the main text and reproduced here for convenience: 
\begin{equation}
s_{l, m}(\vec{x}) = \beta_{l, m}\textrm{sin}\left(\frac{l\pi z}{h}\right)J_0\left(\frac{2j_{0,m}r}{d} \right).
\label{eq:Smode}
\end{equation}
$J_0$ is the zeroth order Bessel function of the first kind and $j_{0,m}$ is the $m^{\textrm{th}}$ root of $J_0$. $h$ is the height of the substrate and $d$ is the diameter of the disk. $\beta_{n, m}$ is a normalization factor so that the total energy of the mode equals $\hbar\omega_{q, m}$, where the mode frequency is given by
\begin{equation}
\omega_{l, m} = \sqrt{\left(\frac{l\pi }{h}\right)^2 v_l^2+\left(\frac{2j_{0,m}}{d}\right)^2 v_t^2 }.
\label{eq:Swnm}
\end{equation}
Here $v_l$ and $v_t$ are the longitudinal and effective transverse sound velocities, respectively. Therefore, we have 
\begin{equation}
\beta_{l, m} = \sqrt{\frac{\hbar\omega_{l, m}}{\pi h c_{33} \displaystyle\int{J_0\left(\frac{2j_{0,m}r}{d} \right)r dr}}}.
\end{equation}

To find the coupling of these modes to the qubit, we can write the mechanical interaction energy as $H = -\int{\sigma(\vec{x}) s(\vec{x})}\: dV = -\int{ c_{33} d_{33}(\vec{x}) E(\vec{x}) s(\vec{x})}\: dV$. Here, $\sigma(\vec{x})$ is the stress, $E(\vec{x})$ is the qubit's electric field profile, and $c_{33}$ and $d_{33}$ are the stiffness and piezoelectric tensor components, respectively. For simplicity, we are only considering the dominant tensor components perpendicular to the surface of the substrate. Quantizing the qubit mode as $E(\vec{x}) (a+a^{\dagger})$ and the phonon mode as $s(\vec{x}) (b+b^{\dagger})$, we can use the rotating wave approximation to equate this interaction energy to the Jaynes-Cummings Hamiltonian $H_{\textrm{int}} = -\hbar g(ab^\dagger+a^\dagger b)$. We then assume that the electric field is a constant $E_0$ throughout the transducer so that we maintain the cylindrical symmetry of the problem. In reality, the electric field is slightly higher in the parts of the AlN that are closer to the edges of the qubit electrode. In addition, by design, the thickness of the AlN is approximately $\lambda_a/2$, where $\lambda_a$ is the acoustic wavelength. Finally, making use of the fact that $d_{33}(\vec{x})$ is a constant $d_0$ inside the AlN disk and zero elsewhere, we find
\begin{equation}
\hbar g_{l,m} = 2c_{33}d_0 E_0 \lambda_a \beta_{l, m} \displaystyle\int_0^{d/2}{J_0\left(\frac{2j_{0,m}r}{d} \right)r dr}
\label{eq:gs}
\end{equation}

\begin{figure}
\begin{center}
\includegraphics[width = 0.9 \textwidth]{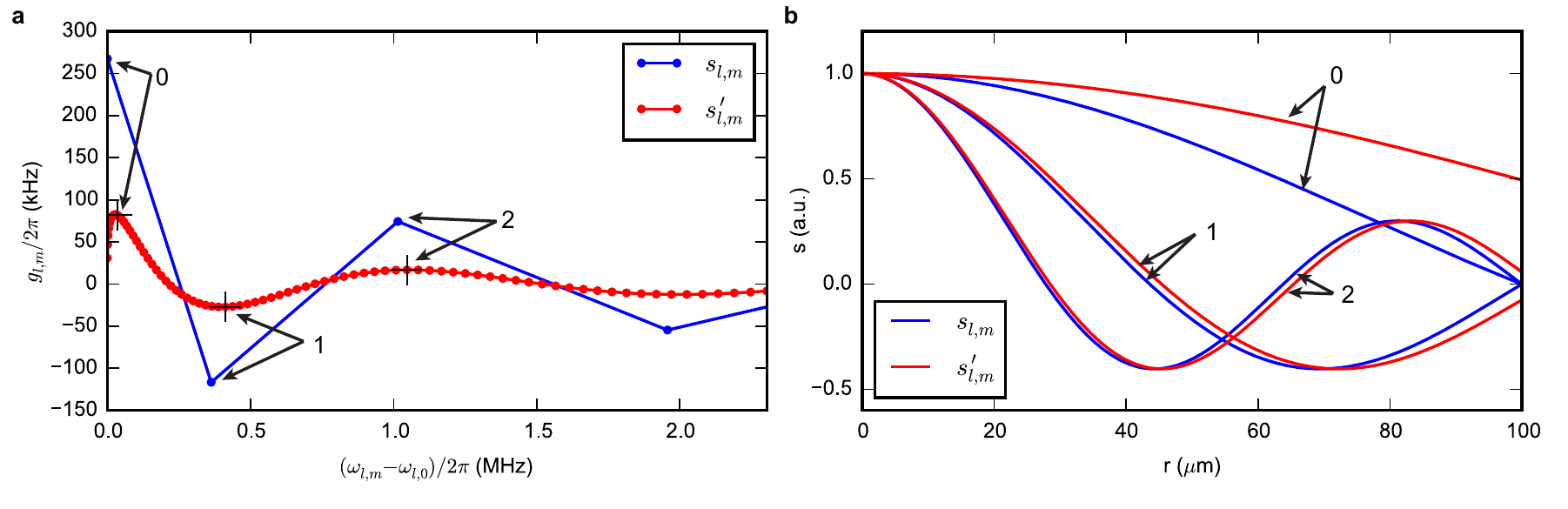}
\caption{\textbf{Phonon modes and qubit coupling a}, Qubit coupling strengths for the $s_{l, m}(\vec{x})$ and $s'_{l, m}(\vec{x})$ modes for $l = 503$ as a function of mode frequency. Arrows and black crosses indicate the modes plotted in \textbf{b}. For the $s_{l, m}(\vec{x})$ modes, $m = 0, 1, 2$ are chosen. For the $s'_{l, m}(\vec{x})$, we chose $m = 9, 33, 53$, which correspond to the local extrema of the coupling strength. \textbf{b}. Comparison of the radial parts of the phonon wavefunctions. By design, the $s_{l, m}(\vec{x})$ modes go to zero at the transverse boundary of the AlN disk, which has a radius of 100 $\mu$m. The $s'_{l, m}(\vec{x})$, instead, go to zero at $r = 2$ mm.}
\label{modes}
\end{center}
\end{figure}

In Figure \ref{modes}a, we show $g_{l,m}$ as a function of $\omega_{l, m}$ for $l = 503$ and $m$ from 0 to 3. Figure \ref{modes}b shows the $m=0, 1, 2$ radial wavefunctions. The estimated value of $g\sim2\pi\times300$ kHz quoted in the main text was obtained using the following constants in equations \ref{eq:wnm} and \ref{eq:gs}: $v_l = 1.11\times10^4$ m/s, $v_t = 8.78\times10^3$ m/s, $c_{33} = 390$ GPa, $d_0 = 1$ pm/V, and $E_0 = 2.9\times 10^{-2}$ V/m. Note that the value of $v_t$ is not simply the transverse sound velocity, but an effective value obtained by fitting the acoustic dispersion surface\cite{SRenninger2016}. The values of $g$ shown in Figure \ref{modes}a include an additional scale factor to match the experimental data, as explained below.

\begin{figure}
\begin{center}
\includegraphics[width = 0.9 \textwidth]{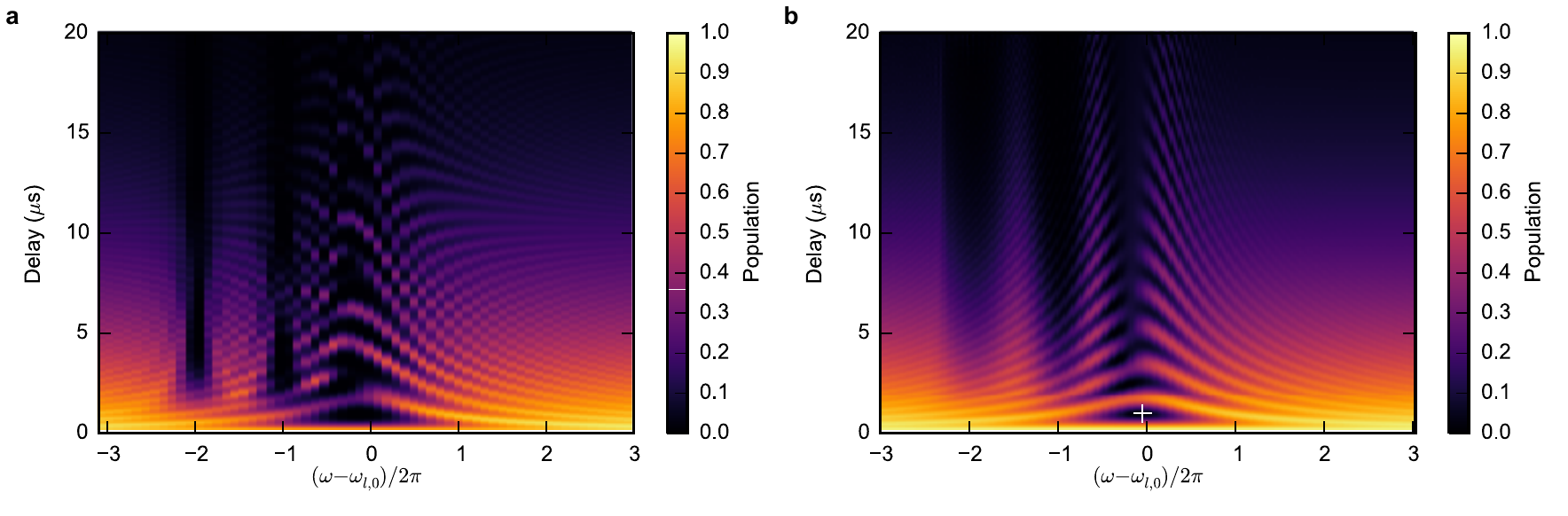}
\caption{\textbf{Simulated Vacuum Rabi oscillations a}, Results of full quantum simulation using the $s_{l, m}(\vec{x})$ modes for $m = 0$ to $3$ and $l = 503$. The horizontal axis shows the qubit detuning from the $m=0$ phonon mode. \textbf{b}, Results of simulations using the $s'_{l, m}(\vec{x})$ modes for $l = 503$, $m = 0$ to $80$ and evolving the equations of motion for the mode amplitudes, as described in the text. White cross indicates the frequency and duration of the swap operation used in calculations of phonon $T_1$ and $T_2$.}
\label{MD}
\end{center}
\end{figure}

We can now take $g_{l,m}$ and $\omega_{l, m}$ as the qubit coupling strengths and frequencies of discrete phonon modes and perform a full quantum mechanical simulation of vacuum Rabi oscillation experiments. As in the experiment, the state is initialized with the qubit in the excited state and all phonon modes in the vacuum state. It then evolves for a delay time according to the Hamiltonian
\begin{equation}
H =\delta_q a^{\dagger}a +\sum_m \delta_{l,m} b^{\dagger}_{l,m}b_{l,m} + g_{l,m}(a^{\dagger}b_{l,m}+ab^{\dagger}_{l,m}),
\label{eq:Ham}
\end{equation}
where we are only summing over $m$ and considering a single $l$ since the longitudinal FSR is large compared to the frequency range we're interested in. $a$ and $b_{l,m}$ are the annihilation operators for the qubit and phonons, respectively. We have used the rotating frame at frequency $\omega_{l,0}$ so that $\delta_q$ and $\delta_{l,m}$ are the detunings of the qubit and phonons from this frequency.
We show the qubit population at the end of this evolution in Figure \ref{MD}a. The simulation were performed using the QuTip python package\cite{SJohansson2012}. To limit computation time, only the $m = 0$ to 3 modes were used. Along with the experimentally measured qubit lifetime, phenomenological phonon lifetimes were introduced for each phonon mode into the Lindblad master equation. We extract a coupling strength of $g_{l,0} = 2\pi\times 260$ kHz from our experimental data by applied an overall scaling to the coupling constants to match the simulated results to the vacuum Rabi data. This is equivalent to multiplying the 	quantity $d_0 c_{33} E_0$ by a factor of 0.85, which is justified since $d_0$ and $c_{33}$ are material constants have not been independently measured for our materials at milikelvin temperatures, and $E_0$ is an approximate value taken from simulations. We see that the simulation results agree reasonably well with the experimental data, indicating that the system can be modeled by a few discrete modes.

The introduction of phenomenological lifetimes for each phonon mode is necessary because we chose a fictitious mode volume from which the energy is lost due to diffraction. Even though the picture of discrete modes with loss rates seems to model the data quite well, energy loss due to diffraction is not a Markovian process and we do not generally expect the time dynamics to follow an exponential decay. For a more realistic model without any additional assumptions, we can instead actually analyze the diffraction of phonons into a larger volume, say a cylinder with radius $a\gg d$. This can be done by finding the coupling of the qubit to the modes $s'_{l, m}(\vec{x})$ of this large cylinder, which have the same wavefunctions and frequencies as given in \ref{eq:gs} and \ref{eq:wnm}, except with $d$ replaced with $a$ everywhere. In Figure \ref{modes}a, we have plotted the coupling strength and frequencies of the modes for a cylinder with $a = 2$ mm. We see that the $s'_{l, m}(\vec{x})$ modes are much more closely spaced in frequencies than the $s_{q, m}(\vec{x})$ modes, as expected from the larger mode volume. However, the shape of the transducer still determines the overall envelope of the coupling strengths as a function of frequency, and we can identify local extrema that correspond to the $s_{q, m}(\vec{x})$ modes. To illustrate this correspondence, we have plotted the wavefunctions of the $s'_{l, m}(\vec{x})$ at these extrema in Figure \ref{modes}b. We see that they resemble the wavefunctions of the $s_{l, m}(\vec{x})$ modes. 

In principle, we can now repeat the simulation of the vacuum Rabi oscillations in the same way as for the $s_{l, m}(\vec{x})$ modes. However, the more realistic picture comes at the cost of more computational complexity. In the same frequency range, instead of four modes, we now have 80, and full quantum mechanical calculation becomes impractical. However, if we start with the qubit in the e state, the Hamiltonian in equation \ref{eq:Ham} only couples the initial state $\ket{1}_q\ket{\textrm{vac}}_p$ to the states $\ket{0}_q\ket{l, m}_p$, where the $q$ subscript denotes the qubit and $p$ denotes the phonon. $\ket{\textrm{vac}}_p$ denotes all phonon modes in the vacuum state, and $\ket{l, m}_p$ denotes the $s'_{l, m}(\vec{x})$ mode in the $n=1$ Fock state and all other modes in the vacuum state. Therefore, to reproduce the vacuum Rabi oscillations, we can just solve the equations of motion for the amplitudes of these states:
\begin{eqnarray}
\dot{c}_q(t) &=& (-\gamma_q+i\delta_q) c_q(t) + \sum_m i g_{l,m} c_{l,m}(t)\\
\dot{c}_{l,m}(t) &=&  i g_{l,m} c_q(t) + i\delta_{l,m} c_{l,m}(t)
\label{eq:emot}
\end{eqnarray}
Note that here, we have included the experimentally measured qubit decay rate, but the phonon modes are lossless.

In Figure \ref{MD}b, we plot $\left|c_q\right|^2$ as a function of interaction time for various $\delta_q$ with $l = 503$. We see that it also approximately produces the main features of the vacuum Rabi data. Unlike Figure \ref{MD}a, these simulation results do not assume discrete phonon modes with phenomenological decay rates. They show that the behavior of our system is well described by a first principles model of phonon propagation and diffraction in a semi-infinite system.

The calculations above are valid for a resonator with two completely flat mirrors. In reality, the AlN disk protrudes above the surface and provides some confinement for the phonons. Therefore, our physical system is likely to be in an intermediate regime between the two models described above, as evidenced by the qualitative change in the phonon spectrum as a function of longitudinal mode number. Neither model completely captures the details the system dynamics, and the more subtle characteristics need to be simulated using more sophisticated methods, especially if we want to use more complicated geometries such as curved surfaces to confine the phonons and simplify the mode structure. We will demonstrate such a method in section \ref{BPS}. First, however, we show how the semi-continuum model can be used to estimate the lifetime of phonons.


\subsection{Phonon lifetime}
In this section, we present a way of estimating the diffraction limited lifetime of the phonons, as measured in the $T_1$ experiment presented in Figure 4a of the main text. In this experiment, we initialize the phonons in some state $s(\vec{x}, \tau=0)$ after a swap operation. In our calculations, we decompose $s(\vec{x}, \tau=0)$ into the $s'_{l, m}(\vec{x})$ basis, with amplitudes $c_{l,m}$ obtained by integrating the equation \ref{eq:emot} for a time $t_{\textrm{swap}}$. Just as in the experiment, we chose a qubit frequency and $t_{\textrm{swap}}$, indicated by a white cross in Figure \ref{MD}b, which most efficiently transfers the qubit state to the phonons. Since the $s'_{l, m}(\vec{x})$ are eigenmodes of the system, we can simply calculate the state of the phonons after diffracting for a time $\tau$ as
\begin{equation}
s(\vec{x}, \tau) = c_{l,m} e^{-i \omega_{l,m}\tau} s'_{l, m}(\vec{x}).
\end{equation}
In the experiment, doing the second swap operation and measuring the state of the qubit corresponds to measuring the overlap of the phonon state $s(\vec{x}, \tau)$ with the original state $s(\vec{x}, \tau=0)$, which we will call $\eta(\tau)$. Specifically, the measured signal is proportional to 
\begin{eqnarray}
\left|\bracket{s(\vec{x}, \tau=0)}{s(\vec{x}, \tau)}\right|^2  &=&\left| \eta(\tau)\right|^2\\
& =& \left| \sum_m{\left|c_{l,m}\right|^2 e^{i \omega_{l,m}\tau}}\right|^2
\end{eqnarray}

\begin{figure}
\begin{center}
\includegraphics[width = 0.5 \textwidth]{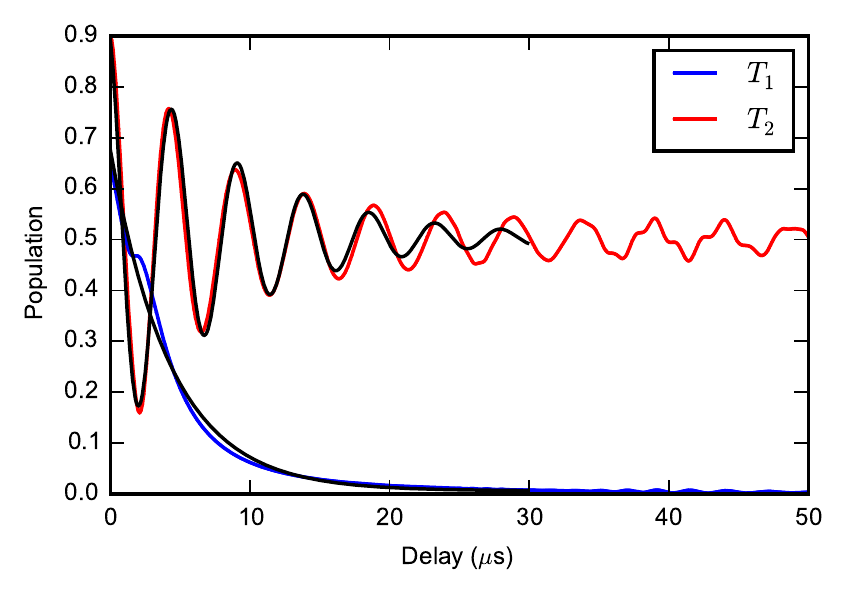}
\caption{\textbf{Simulated phonon $T_1$ and $T_2$.} Black lines are exponential and sinusoidally modulated exponential fits to a part of the simulated $T_1$ and $T_2$ data, respectively. Here we have used the same artificial detuning of $\Omega =2\pi\times 200$ kHz as in the experiment. The fits are not ideal due to the effects of the finite simulated mode volume, which result in reflections at long times. We estimate $T_1 \sim$ 4 $\mu$s and $T_2 \sim$ 7$\mu$s from the fit.}
\label{fig:T1T2Sim}
\end{center}
\end{figure}
Similarly, one can find that the measured $T_2$ signal has the form $\left(1+Re\left[e^{i\Omega\tau} \eta(\tau)\right]\right)/2$. We plot the simulated $T_1$ and $T_2$ results in Figure \ref{fig:T1T2Sim}. We see that the simulation qualitatively reproduces the features of the data in Figure 4 of the main text, with a few deviations. First, we see the effects of the finite mode volume at long times, when the simulated wavefunction starts to reflect back from the boundaries. Second, the timescales are a few microseconds, which is on the same order as but somewhat shorter than the experimentally measured values. This could be due to the confinement of the modes by the AlN in the physical system, which is not included in the simulation. Finally, the initial populations in the simulated data are higher because qubit decay during the second swap pulse is not included.

\section{Beam propagation simulations: Finding acoustic modes with AlN}
\label{BPS}

\begin{figure}
\begin{center}
\includegraphics[width = 0.5 \textwidth]{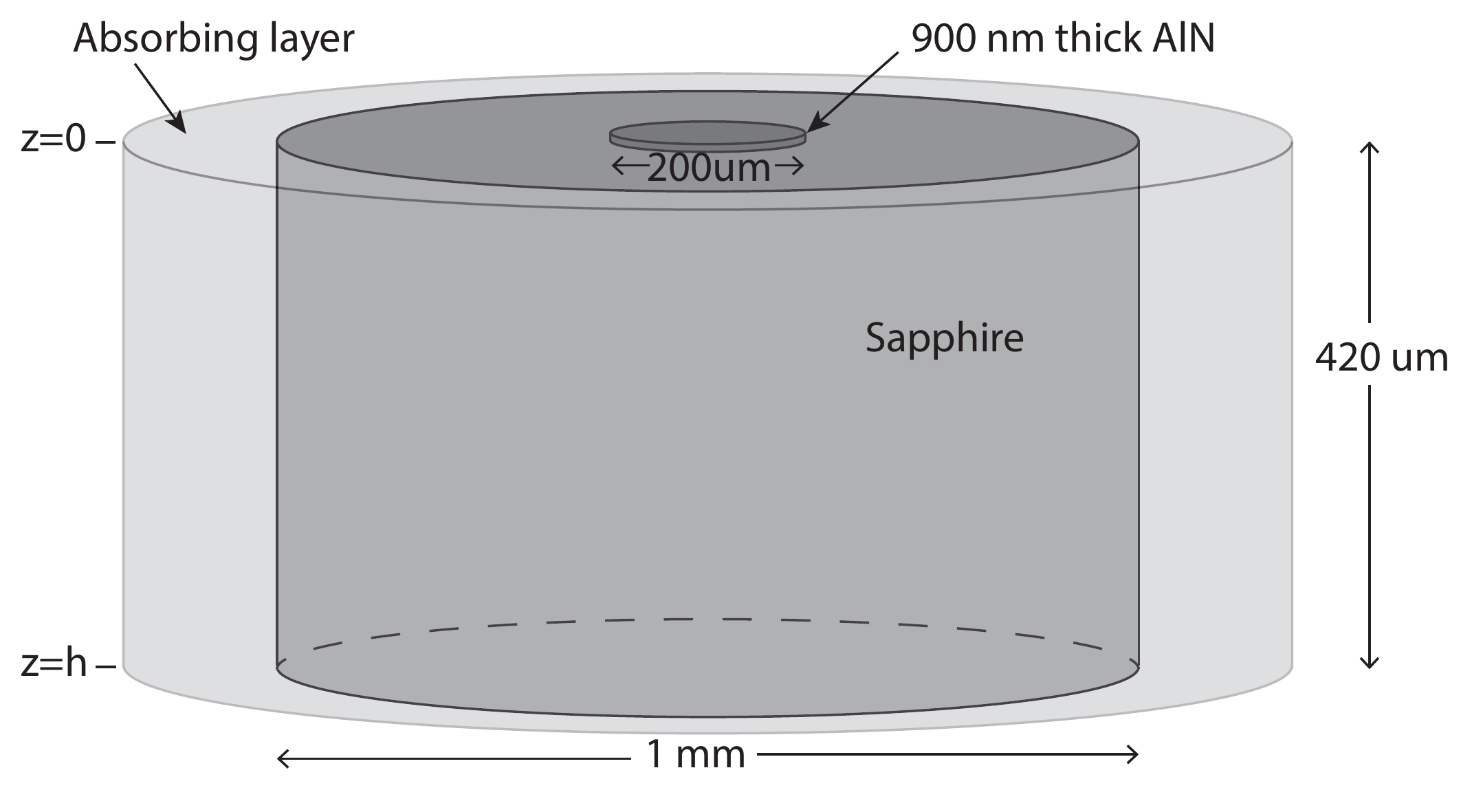}
\caption{\textbf{Simulation geometry including the AlN disk.} We consider a cylindrical sapphire substrate with two flat faces. The diameter of the cylindrical region is 1 mm and the height is 420 $\mu$m. An AlN disk of thickness 900 nm and diameter 200 $\mu$m protrudes from the center of one face. An absorbing layer is added to mitigate unwanted reflection from the transverse boundaries and simulate lossy boundaries of the sapphire. We use 11110 m/s and 6056 m/s as the longitudinal and transverse velocities for sapphire, respectively. We use 11008 m/s for the longitudinal velocity of sound in AlN.}
\label{fig:SimRegion}
\end{center}
\end{figure}

In this section, we use acoustic beam propagation and the method of resonant acoustic excitation to determine acoustic modes of our system with the effects of AlN disk included\cite{SRenninger2016}. We were motivated to develop this simulation technique not just to understand any additional effects of the AlN, but also to develop a method to find the acoustic modes of an arbitrary geometry (such as a plano-convex resonator), crystalline material, and drive region for future device improvements. 

\begin{figure}
\begin{center}
\includegraphics[width = 0.5 \textwidth]{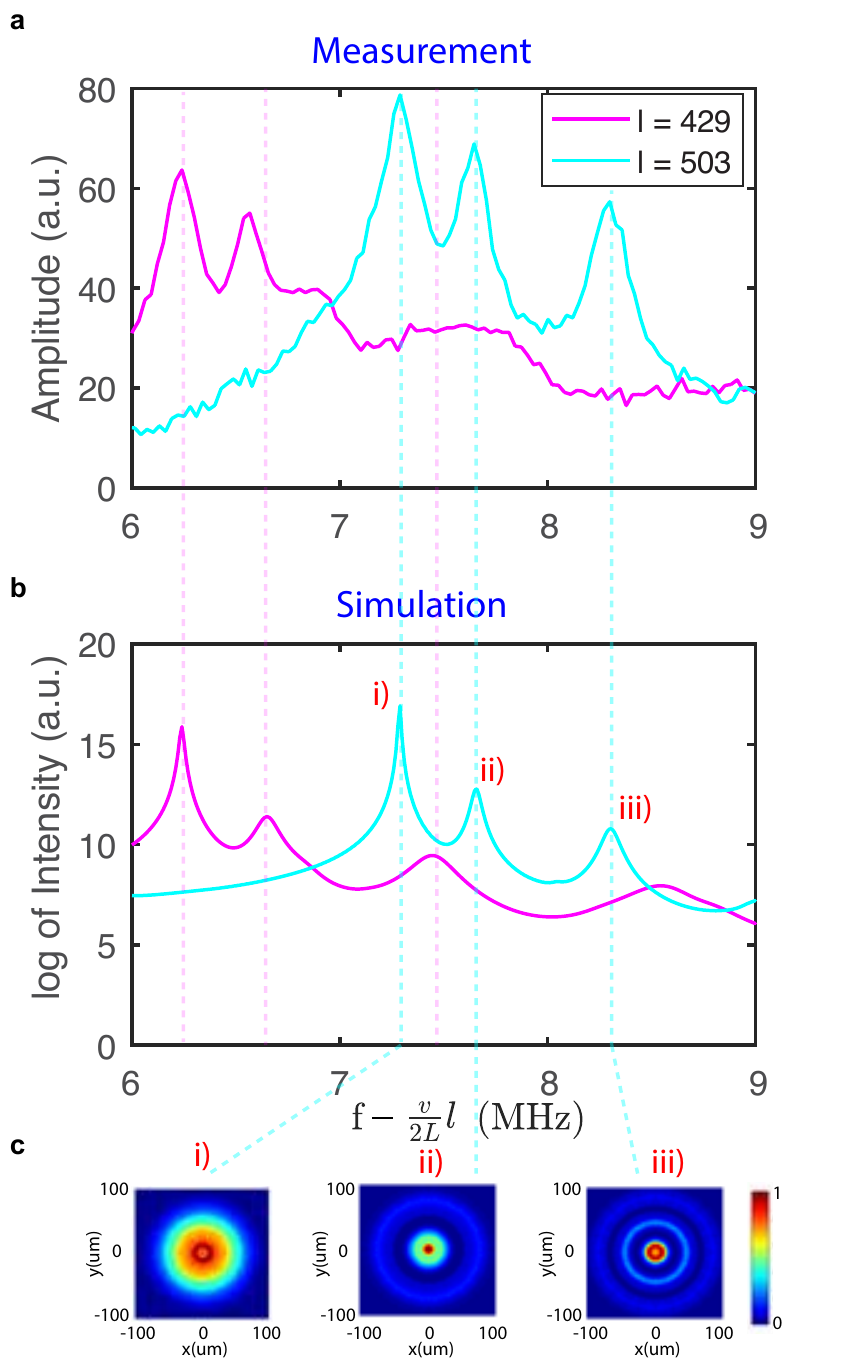}
\caption{\textbf{Resonant frequencies and mode profiles including the effect of AlN.} \textbf{a.} Maximum value plot taken from the spectroscopy data for modes around $l=429$ and $l = 503$. The amplitude axis is inverted for ease of comparison with simulation results. \textbf{b.} Total acoustic intensity integrated over the z= 0 plane showing the resonant frequencies from the beam propagation simulation for $l=429$ and $l=503$ modes. \textbf{c.} Acoustic intensity profiles of the resonant acoustic modes around $l=503$. These mode profiles are cross-sections of the 3D acoustic modes taken at $z=0$. }
\label{fig:freqSweep}
\end{center}
\end{figure}

The simulation region we used is shown in Figure \ref{fig:SimRegion}. To find the resonant thickness modes of this system, we start with an initial field distribution that is longitudinally polarized at $z= 0$. Since we assume a constant piezoelectric stress (or constant drive) in the AlN region, we choose an initial acoustic field that is constant in the AlN region and zero elsewhere. We then use Fourier beam propagation, analogous to the technique used in optics\cite{Fox1968}, to compute the acoustic field after traveling through a distance of $2h$ in the crystal. An extra phase after each round trip is applied to account for the AlN disk. This process is repeated for multiple round trips inside the crystal. A complex interferometric sum of fields at $z= 0$ after each roundtrip is also recorded. As we sweep the frequency of the drive and repeat the beam propagation, the total intensity of this complex sum of fields resonantly builds up for some frequencies, as shown in Figure \ref{fig:freqSweep}. These resonant frequencies correspond to the acoustic modes whose phase and amplitude profile remain unchanged after each round-trip.

Once the resonance frequency $\Omega_m$ is found, the initial field driven at $\Omega_m$ is propagated over multiple roundtrips and a complex sum of displacement fields is computed. This complex sum is then used as an input to another beam propagation over multiple round trips. After several iterations, the complex sum does not change and converges to the mode profiles of the standing wave acoustic mode at frequency $\Omega_m$. Using the resonant frequencies and mode profiles, we can then calculate the quantities of interest to our system, such as coupling rates to the qubit.

We compare the features in the maximum value plot obtained from the spectroscopy data (See Figure \ref{FSRs}) around the $l = 429$ and $l = 503$ modes with the resonant frequencies from the simulation in Figure \ref{fig:freqSweep}. Since the absolute frequencies of simulated modes depend on the longitudinal sound velocity, which is not precisely known, we have lined up the $m = 0$ modes in the maximum value plot for both $l = 429$ and $l=503$ modes with their respective frequency values from the simulation. This allows us to compare the relative spacings of the higher order transverse modes. The dotted lines between Figure \ref{fig:freqSweep}b and \ref{fig:freqSweep}c suggests that the relative spacing between the higher order modes in the simulation matches reasonably well with the measurement. The simulations indicate that at lower longitudinal mode number $l$ (corresponding to smaller phonon frequencies), the separation between the higher order transverse modes increases as observed in the experiment. It can be observed from the data that the ratios of separations also change, and this is qualitatively reproduced by the simulations. We point out that the the maximum value plots are not a direct measurement of the total field intensity of the acoustic modes and depends on more complex aspects of the system such as the coupling to the qubit. Discrepancies between the data and simulation can be attributed to this, along with other factors such as non-uniformities in the electric field and uncertainties in the AlN thickness. Therefore we simply aim to compare the frequency spacings of the modes and verify that our understanding of the experimentally observed mode structure is valid with the inclusion of the AlN. 

Once the resonant mode frequencies are found, we computed acoustic intensity profiles for modes around $l = 503$ at $z=0$ (See Figure \ref{fig:freqSweep} c)). These acoustic mode intensity plots show that most of the acoustic energy for these modes is confined within the AlN region. These modes nonetheless extend all throughout the simulation domain and suffer absorption losses at the simulation boundaries. The leaky nature of these modes is also evident in the finite width of the resonant spectrum in our simulation. 

For future device improvements, we will use these simulation techniques to design resonators with shaped boundaries, such as a plano-convex geometry. This should allow for lateral confinement of thickness acoustic modes and result in a dramatic increase in phonon coherences. Additionally, these simulations will help us identify the correct shape for the AlN drive region so that we can tailor the coupling to a single phonon mode.

\end{document}